\documentclass[12pt]{iop/iopart}
\usepackage[utf8]{inputenc}
\usepackage{iop/iopams}
\usepackage{graphicx}
\usepackage{latexsym}
\usepackage{comment}
\usepackage{bbm}
\usepackage{bm}

\newcommand{\conj}[1]{\bar{#1}}
\newcommand{\vb}[1]{\bm{\mathrm{#1}}}

\newcommand{\abs}[1]{|#1|}
\renewcommand{\>}{\rangle}
\newcommand{\<}{\langle}
\renewcommand{\exp}[1]{{\rm exp}\left\{#1\right\}}
\newcommand{\com}[2]{\left[#1,#2\right]}
\renewcommand{\i}{\rmi}

\begin{document}

\title{Quantum Interference on the Non-Commutative Plane and the Quantum-to-Classical Transition}  
\author{IB Pittaway$^{a}$ and FG Scholtz$^{a,b}$ }
\address{$^a$Department of Physics, Stellenbosch University,
Stellenbosch 7602, South Africa}
\address{$^b$Institute of Theoretical Physics, 
Stellenbosch University, Stellenbosch 7602, South Africa\\}

\begin{abstract}
We explore a possible link between the structure of space at short length scales and the emergence of classical phenomena at macroscopic scales.  To this end we adopt the paradigm of non-commutative space at short length scales and explicitly compute the outcomes of a double slit experiment and a von Neumann measurement in the non-commutative plane.  A very consistent picture of a continuous quantum-to-classical transition emerges.  The mechanism driving this transition is standard decoherence, but here the ``environment'' arises quite naturally from the tensor product structure of the non-commutative quantum Hilbert space.  The double slit calculation enables us to establish  a lower bound on the non-commutative parameter for this transition to become effective at particle numbers of the order of Avogadro's number.  Similarly, the result of the von Neumann measurement establishes a criterium involving the non-commutative parameter, apparatus size and coupling between system and apparatus for classicality to emerge.
\end{abstract}


\maketitle


\section{Introduction}\label{sec:intro}
There are two open questions in modern physics that have generated enormous debate and are still contentious. 
The first is the unification of gravity and quantum mechanics, which is intimately tied to the structure of space-time at short length scales and the emergence from it of space-time as we perceive it at long length scales (see for example \cite{seib}). 
One such structure is that of non-commutative space-time, which has attracted considerable attention and is supported by compelling arguments presented in \cite{dop}. It is also further supported by its natural emergence in certain string theories \cite{wit}.
The second open question is that of the transition between the quantum and classical paradigms, often referred to as the measurement problem, which is predicated on the problem of macroscopic objectivity or realism (see for example \cite{legget}). 

These problems are almost always addressed as distinct, unrelated problems, however, there have been arguments by Penrose \cite{pen} that suggest gravity might play a role in state reduction.
With this in mind, we explore here a possible connection between the structure of space-time at short length scales and the suppression of quantum effects at larger length scales.

We adopt non-commutative space as our paradigm for space at short length scales and explicitly compute, within this setting, the outcome of a double-slit experiment, as well as the dynamics of a bipartite system during a von Neumann style measurement procedure.  
Here we restrict ourselves to the simplest possible case: that of constant commutation relations in two dimensions, generally referred to as the non-commutative plane. This restriction is chosen for the comparative simplicity it affords us during this investigation.
Remarkably, it turns out that the usual interference term in the double-slit experiment, as well as the quantum correlations between measurement results during von Neuman measurements, are both suppressed in macroscopic systems. It would seem that a non-commutative space-time might lead to a mass dependent suppression of the quantum properties of the centre of mass motion of a macroscopic object. 
These promising results then offer motivation for further expansion onto the more mathematically complex construction of three dimensional non-commutative fuzzy space.

A consistent formulation of quantum mechanics on the non-commutative plane has been done in \cite{scholtz1} where it was shown to be standard quantum mechanics implemented on a Hilbert-space of Hilbert-Schmidt operators. The only novelty is that, due to the non-commutativity of space, position measurement must now be implemented as a weak measurement through a positive operator-valued measure (POVM). This construction has also been generalised to three dimensional fuzzy space \cite{scholtz2}, where one adopts $SU(2)$ commutation relations for the non-commutative coordinates, which avoids the breaking of rotational symmetry encountered with constant commutation relations.

This paper is organized as follows: Section \ref{sec:formalism} gives a brief overview of quantum mechanics on the non-commutative plane and develops the notion of a position measurement in more detail. Section \ref{sec:doubleslit} computes the outcome of a double-slit experiment in the non-commutative plane and establishes that in principle interference is suppressed at high momentum.  Section \ref{sec:sternGerlach} computes the outcome of a von Neumann measurement in which a spin system is coupled to the centre of mass momentum of a pointer. This demonstrates that in principle the density matrix becomes an effectively diagonal mixed state density matrix at length scales above the square root of the non-commutative parameter.  Section \ref{sec:macroLimit} takes into account the role of macroscopic effects in the double-slit experiment and the von Neumann measurement and establishes that in the macroscopic limit all quantum interference effects are strongly suppressed.  Section \ref{sec:discussion} closes with a summary and discussion.                   

\section{Quantum mechanics on the non-commutative plane and position measurement}\label{sec:formalism}

As the quantum theory on the non-commutative plane \cite{scholtz1} is central to our computation, we briefly review it. 
The foundation of the theory is based on the introduction of a non-commutative configuration space, which we denote by ${\cal H}_c$, to replace the standard commutative configuration space (i.e. ${\mathbbm R}^2$). 

We assume that the coordinate algebra in two dimensions is given by \cite{scholtz1}
\begin{equation}\label{ncom2d}
    \com{\hat{x}}{\hat{y}} = \i \theta.
\end{equation}
Here $\theta$ is a constant with the dimension of a length squared, which we take without loss of generality to be positive. 
We use a hat to denote that  the coordinates are operators, but will drop it where no confusion can arise.
To develop a quantum theory we require a unitary representation of this algebra on the configuration Hilbert space ${\cal H}_c$.
Noting that $b=\frac{1}{\sqrt{2\theta}}(\hat{x}+\i\hat{y})$ and $b^\dagger=\frac{1}{\sqrt{2\theta}}(\hat{x}-\i\hat{y})$ are standard creation and annihilation operators, an obvious candidate for ${\cal H}_c$ is the Fock space for a single oscillator \cite{scholtz1}. This makes further sense if one notes that the radius operator is $\hat{r}^2=\hat{x}^2+\hat{y}^2=\theta (b^\dagger b+1)$. Thus each value of the quantized radius appears exactly once in this representation and in this sense the two-dimensional plane is completely covered once.

This configuration space is not the one in which the quantum states of the system are to be represented. For this we need the quantum Hilbert space which we denote by ${\cal H}_q$. This space carries the representation of the full non-commutative Heisenberg algebra.
In the two dimensional case, this is the Hilbert space of all Hilbert-Schmidt operators on ${\cal H}_c$ that are generated by the non-commutative coordinates, making ${\cal H}_q$ isomorphic to the tensor product ${\cal H}_c\otimes{\cal H}_c^*$. Here ${\cal H}_c^*$ is the dual of the configuration Hilbert space. 
Generally, the quantum Hilbert space is only a subspace of all Hilbert-Schmidt operators generated by the non-commutative coordinates \cite{scholtz2}, as is the case in three dimensional fuzzy space. 

To avoid confusion, we denote states in ${\cal H}_c$ by $|\cdot\>$ and states in ${\cal H}_q$ by $|\cdot)$. A general element of ${\cal H}_q$ then has the form $|\psi)\equiv\sum_{n,m} \psi_{n,m} |n\>\< m|$ with $\sum_{n,m} |\psi_{n,m}|^2<\infty$. Note that the states $|n,m)=|n\>\<m|$ form a complete orthonormal basis in ${\cal H}_q$.  
The inner product on ${\cal H}_q$ is then $(\phi|\psi)={\rm tr}(\phi^\dagger\psi)$ where ${\rm tr}$ denotes the trace over ${\cal H}_c$, We use ${\rm Tr}$ to denote the trace over ${\cal H}_q$, while $\dagger$ and $\ddagger$ are used to denote hermitian conjugation on ${\cal H}_c$ and ${\cal H}_q$ respectively.

The construction of the quantum theory now proceeds as normal: one introduces observables as hermitian operators acting on ${\cal H}_q$ with the standard probabilistic interpretation. To distinguish these observables from operators on ${\cal H}_c$, we commonly denote them by capitals. 
We proceed to define the most important observables. The quantum position operators are defined through left action,
\begin{equation}\label{qpos}
    X|\psi)=\left|\hat{x} \psi\right),\quad Y|\psi)=|\hat{y} \psi),\quad R^2=X^2+Y^2,
\end{equation}
and the quantum momentum operators are defined by the adjoint action, i.e. its action on a generic element  $|\psi)$  in ${\cal H}_q$ is given by
\begin{equation}\label{momadj}
    P_i|\psi)= \left|\frac{\hbar}{\theta}\epsilon_{ij}\left[\hat{x}_j,\psi\right]\right),
\end{equation}
with $\epsilon_{ij}$ the completely anti-symmetric tensor. It turns out to be convenient to rather consider the complex momentum $P=P_x+\i P_y$ and its conjugate $P^\ddagger$ for which the actions are
\begin{equation}
    P|\psi)=\left|-\i\hbar\sqrt{\frac{2}{\theta}}[b,\psi]\right),\quad P^\ddagger|\psi)=\left|\i\hbar\sqrt{\frac{2}{\theta}}[b^\dagger,\psi]\right).
\end{equation}
Note that momentum involves a left and right multiplication, while position only involves left multiplication. The choice of defining position by left multiplication is not unique and a matter of convenience in that any other choice is related by a unitary transformation. 
Finally, the Hamiltonian is given by \cite{scholtz1}
\begin{equation}\label{doubleSplitH}
    H=\frac{P^\ddagger{P}}{2m}+V(X,Y), \qquad V(X,Y)^\ddagger=V(X,Y).
\end{equation}
Another useful observable is the angular momentum, which acts as follows:
\begin{equation}
L|\psi)=|\hbar[b^\dagger b,\psi]).
\end{equation}
If the potential is a function of $\hat{R}$ only, this operator commutes with the Hamiltonian and is a conserved quantity.

For a free particle with $V=0$ the Hamiltonian eigenvalue equation can easily be solved \cite{scholtz1}. The eigenstates are given by non-commutative plane waves 
\begin{equation}\label{pw}
    |{\bf p})
    =\left|\rme^{\i{\bf p}\cdot\hat{\bf x}/\hbar}\right)
    =\left|\rme^{-\frac{\phi^2}{2}\abs{p}^2} \rme^{\i\phi pb^\dagger} \rme^{\i\phi \conj{p}b}\right).
\end{equation}
Here $\phi^2\equiv\frac{\theta}{2\hbar^2}$ and $p=p_x+ip_y$, with $\bar{p}$ its complex conjugate. For later convenience the states have been normalised such that $({\bf p}^\prime|{\bf p})=\frac{2\pi\hbar^2}{\theta}\delta\left({\bf p}^\prime-{\bf p}\right)$. The corresponding energy eigenvalue is $E=\frac{{\bf p}^2}{2m}$.

With this, we have a consistent interpretational framework of standard quantum mechanics, the only exception being that of position measurement. As the coordinates are no longer commuting, we cannot interpret position measurements in the context of Projective Valued Measures (PVMs) as this would require simultaneous eigenstates of position. We are instead required to weaken the notion of a position eigenstates to that of minimal uncertainty states that give the best possible localisation of a particle in position. In this case, these states are very well known to be the standard normalised Glauber coherent states, \cite{klauder}
\begin{equation}
    |z\>=\rme^{-|z|^2/2}\rme^{z b^\dagger}|0\>, \qquad b|z\>=z|z\>, \qquad \int \frac{d\bar{z}dz}{\pi}|z\>\<z|=\mathbbm{1}_c,
\end{equation}
which represent the best approximation to a position eigenstate or point in the non-commutative plane. In this sense, $z$ must then be interpreted as a dimensionless complex coordinate on the plane, $z=\frac{1}{\sqrt{2\theta}}(x+\i y)$, as is clear from the expectation values of $x=\< z|\hat{x}|z\>=\sqrt{2\theta}{\rm Re}{z}$ and $y=\< z|\hat{y}|z\>=\sqrt{2\theta}{\rm Im}{z}$. 

These are still classical configuration space states and equivalents must be found at the quantum level. To this end, consider the quantum position operators acting on ${\cal H}_q$ as defined in (\ref{qpos}). Following from the nature of $z$, it is more convenient to consider the quantum analogues of $b$ and $b^\dagger$ defined by
\begin{equation}
    B=\frac{1}{\sqrt{2\theta}}\left(\hat{X}+\i\hat{Y}\right), \qquad B^\ddagger=\frac{1}{\sqrt{2\theta}}\left(\hat{X}-\i\hat{Y}\right).
\end{equation}
These satisfy $[B,B^\ddagger]=1$ and their action on elements of ${\cal H}_q$ can be deduced from (\ref{qpos}):
\begin{equation}
    B|\psi)=|b\psi),\qquad B^\ddagger|\psi)=|b^\dagger\psi).
\end{equation}
We note that elements of ${\cal H}_q$ of the form $|z,n)=|z\>\<n|$ satisfy 
\begin{equation}
    B|z,n)=z|z,n)\quad\forall n,
\end{equation}
and are the coherent states in ${\cal H}_q$. Note that $n$ is arbitrary due to the left action of the position operators. They only measure the left-hand sector of the quantum state.

It is now natural to define the position projection operators on ${\cal H}_q$ as
\begin{equation}\label{POVM}
    \Pi_z=\sum_{n=0}^\infty |z,n)(z,n|.
\end{equation}
These are positive, non-orthogonal operators that resolve the identity on ${\cal H}_q$,
\begin{equation}\label{q_identity}
   \frac{1}{\pi} \int \rmd\conj{z}\rmd z\:\Pi_z=\mathbbm{1}_q,
\end{equation}
and as such constitutes a positive operator-valued measure (POVM). A precise, probabilistic interpretation can now be given to a non-commutative position measurement. Let $\rho_q$ be a density matrix describing the state of the non-commutative quantum system, i.e. it is a non-negative, hermitian, trace one operator acting on ${\cal H}_q$, then the probability density that a position measurement yields the result $(x,y)$, with $x=\sqrt{2\theta}{\rm Re}{z}$ and $y=\sqrt{2\theta}{\rm Im}{z}$, is given by
\begin{equation}\label{prob}
    P(x,y)={\rm Tr}(\Pi_z\rho_q).
\end{equation}
Note that the trace here is over ${\cal H}_q$. For a pure state density matrix $\rho_q=|\psi)(\psi|$, this yields
\begin{equation}\label{prob1}\eqalign{
    P(x,y)&=\sum_{n=0}^\infty(z,n|\psi)(\psi|z,n)=\sum_{n=0}^\infty\< z|\psi|n\>\< n|\psi^\dagger|z\>
    \\
    &=\< z|\psi\psi^\dagger|z\>={\rm tr}(\pi_z\rho_c),
}\end{equation}
where $\pi_z=|z\>\<z|$ and $\rho_c=\psi\psi^\dagger$. Note that $\pi_z$ and $\rho_c$ act on ${\cal H}_c$. It is simple to verify that $\pi_z$ constitutes a POVM on ${\cal H}_c$, while $\rho_c$ is a positive, hermitian, trace one (assuming $(\psi|\psi)=1$) operator and thus a valid, but not necessarily pure, density matrix on ${\cal H}_c$. 
This allows for an interesting alternative interpretation of position measurement as a weak measurement on a generally mixed state in ${\cal H}_c$ alone. 

It is quite simple to understand how this situation can arise:  The quantum Hilbert space ${\cal H}_q$ is isomorphic to a tensor product space ${\cal H}_c\otimes{\cal H}_c^*$, where ${\cal H}_c^*$ is the dual of ${\cal H}_c$. Since the quantum position operator acts by left multiplication, it only acts on one sector ${\cal H}_c$ of the quantum Hilbert space. It therefore corresponds to a local measurement that cannot yield any information on the other sector ${\cal H}_c^*$.  
Under these conditions, as is well-known \cite{schloss1,schloss2,zurek}, a partial trace over the unobserved sector can be performed and a reduced density matrix, in this case $\rho_c$, can be found. This density matrix is generally not a pure state density matrix, i.e. $\rho_c^2\ne\rho_c$. It is instead what is normally referred to as an improper mixed state, i.e. it is not a proper statistical mixture of pure states, but is rather a mixed state derived from a pure state density matrix on the full quantum Hilbert space and as such still encodes quantum information.

Position measurements of a quantum state are then singled out from other types of measurements. They are local measurements of the left-hand sector only, in contrast to momentum which acts on both through its adjoint action. An observer that has access to only position observables would be incapable of measuring the right-hand sector. 

To emphasizes this local nature of an observable, we introduce a new notation. We denote by $O_L$ observables acting on the left of the operator valued wave-function, and similarly $O_R$ an observable that acts on the right,
\begin{equation}
    O_L|\psi)=|o\psi),\qquad O_R|\psi)=|\psi o).
\end{equation}
This can also be expressed in tensorial notation as $O_L=o\otimes\mathbbm{ 1}_c^*$ and $O_R=\mathbbm{1}_c\otimes o$. 
In this notation, we note from (\ref{qpos}) and (\ref{momadj}) that position and momentum operators become
\begin{eqnarray}
    X_i&=&{X_i}_L,
    \\\label{momLR}
    P_i&=&\frac{\hbar}{\theta}\epsilon_{ij}\left({X_j}_L-{X_j}_R\right),
\end{eqnarray}
with additional commutation relations
\begin{eqnarray}
    \com{{X_i}_L}{{X_j}_L} &=& -\com{{X_i}_R}{{X_j}_R} = \i\epsilon_{ij}\theta,
    \\
    \com{{X_i}_L}{{X_j}_R} &=& 0,\quad\forall i,j.
\end{eqnarray}
Naturally, the notion of left and right acting observables can only be applied to observables on the quantum Hilbert space.

By considering the ensemble average of a left acting observable on an arbitrary density operator $\rho=\sum_{\alpha\beta}\rho_{\alpha\beta}|\psi_\alpha)(\psi_\beta|$, we find that
\begin{equation}\eqalign{
    \< O_L\>_q &= {\rm Tr}(O_L\rho)
    =\sum_{\alpha\beta}\rho_{\alpha\beta}(\psi_\beta |O_L|\psi_\alpha)
    \\&=
    \sum_{\alpha\beta} \rho_{\alpha\beta}(\psi_\beta|o\psi_\alpha)
    \\&=
    \sum_ {\alpha\beta} \rho_{\alpha\beta}{\rm tr}(\psi_\beta^\dagger o\psi_\alpha)
    \\&=
    {\rm tr}(o\rho_c) = \<o\>_c,
}\end{equation}
where $\rho_c=\sum_{\alpha\beta}\rho_{\alpha\beta}\psi_\alpha\psi^\dagger_\beta$.  Here we see the equivalence between the statistical expectation value of the quantum operator $O_L$ on the quantum space and that of the classical operator $o$ on the configuration space, regardless of the purity of the quantum state density matrix $\rho$.

The appropriate interpretation of this result, as explained in \cite{schloss2,bub}, is that a measurement of observable $o$ on the reduced density matrix $\rho_c$ in the configuration space ${\cal H}_c$, yields exactly the same statistics as a quantum measurement of a left acting observable $O_L$ on the full density matrix $\rho=|\psi)(\psi|$ in the quantum Hilbert space ${\mathcal H}_q$.
An observer restricted to the measurement of left acting (position) observables only would have no perception or knowledge of the right hand side. 
As such, the quantum state density matrix would be statistically indistinguishable to that of the reduced density matrix in the configuration space. 
Since the reduced density matrix $\rho_c$ is generally an improper mixed state, regardless of the purity of the initial quantum state, this indistinguishability of the statistics would imply that an observer limited to these left acting (position) measurements alone cannot distinguish between a pure or mixed quantum state.
Note that the same is true for right acting observables with the reduced density matrix $\rho_c=\sum_{\alpha\beta}\rho_{\alpha\beta}\psi^\dagger_\alpha\psi_\beta$.

An observer that wants to distinguish between the two situations above, i.e. between measurements on the reduced configuration space density matrix alone as opposed to the full quantum density matrix, needs access to observables that generally mix the left and right sectors.  Such a general observable is of the form
\begin{equation}\label{genob}
    O=\sum_{n,m,i,j}c_{n,m,i,j}{X_i}^n_L{X_j}^m_R.
\end{equation}
From (\ref{momLR}) we note that we can also write
\begin{equation}
    {X_i}_R={X_i}_L+\frac{\theta}{\hbar}\epsilon_{ij}P_j.
\end{equation}
Introducing this into (\ref{genob}), we find that terms that probe the right-hand sector then contain the factor $\frac{\theta}{\hbar}$ whose power $m$ corresponds to the power of the momentum operator. Therefore, unless the pure state density matrix $\rho$ admixes very high momentum states with momentum of the magnitude $p>\frac{\lambda\hbar}{\theta}$, with $\lambda$ the length scale on which the position measurement is being done (units of $x_i$), any terms with high powers $m$ of the momentum operator in the operator's expectation value $\<O\>$ will be severely suppressed. This means that an observer carrying out measurements at low energy or momentum will, from an operational point of view, only have access to observables involving low powers of the momentum operator. To appreciate the orders of magnitude involved here, note that $\frac{\lambda\hbar}{\theta}\sim10^{36}\;{\rm kg.m.s}^{-1}$ if $\theta \sim l_p^2$ and $\lambda\sim{\rm m}$.
However, to probe the right-hand sector and distinguish between a measurement on the full quantum density matrix as opposed to the reduced density matrix, the observer needs access to a complete set of observables including those involving arbitrarily high powers of momentum. 
When an observer measures a system described by a low energy state, any terms in the observable of high order momentum become inaccessible. Operationally then, an observer cannot distinguish between pure and mixed state descriptions as information from the right-hand sector is not available.
This also singles out position, momentum, angular momentum and energy as the preferred observables at low energy.

One cannot fail to notice the analogy with the environmental decoherence program \cite{schloss1,zurek}, which aims to contextualise the quantum-to-classical transition. In that setup a quantum system being measured is coupled to an environment. Upon measurement of system observables only (local measurement), the environmental degrees of freedom can be traced over to yield a reduced system density matrix, which is also generally an improper mixed state. Here, the comparison can be made in which the Hilbert space of the measured system is analogous to ${\cal H}_c$ and the environment is analogous to that of ${\cal H}_c^*$, with the momentum operator acting to couple these spaces. Indeed, one may view the non-commutative quantum setup described above as a natural realisation of the decoherence paradigm. As is well known, environmental decoherence generally leads to the suppression of interference terms, which gives rise to an emerging classical behaviour \cite{schloss1,zurek}. One may therefore expect a similar scenario to play out here, which we explore next through two specific examples, namely, a double slit experiment and a von Neumann type measurement.

\section{The Double Split Experiment}\label{sec:doubleslit}
\begin{figure}
    \centering
  \includegraphics[width=0.2\textwidth]{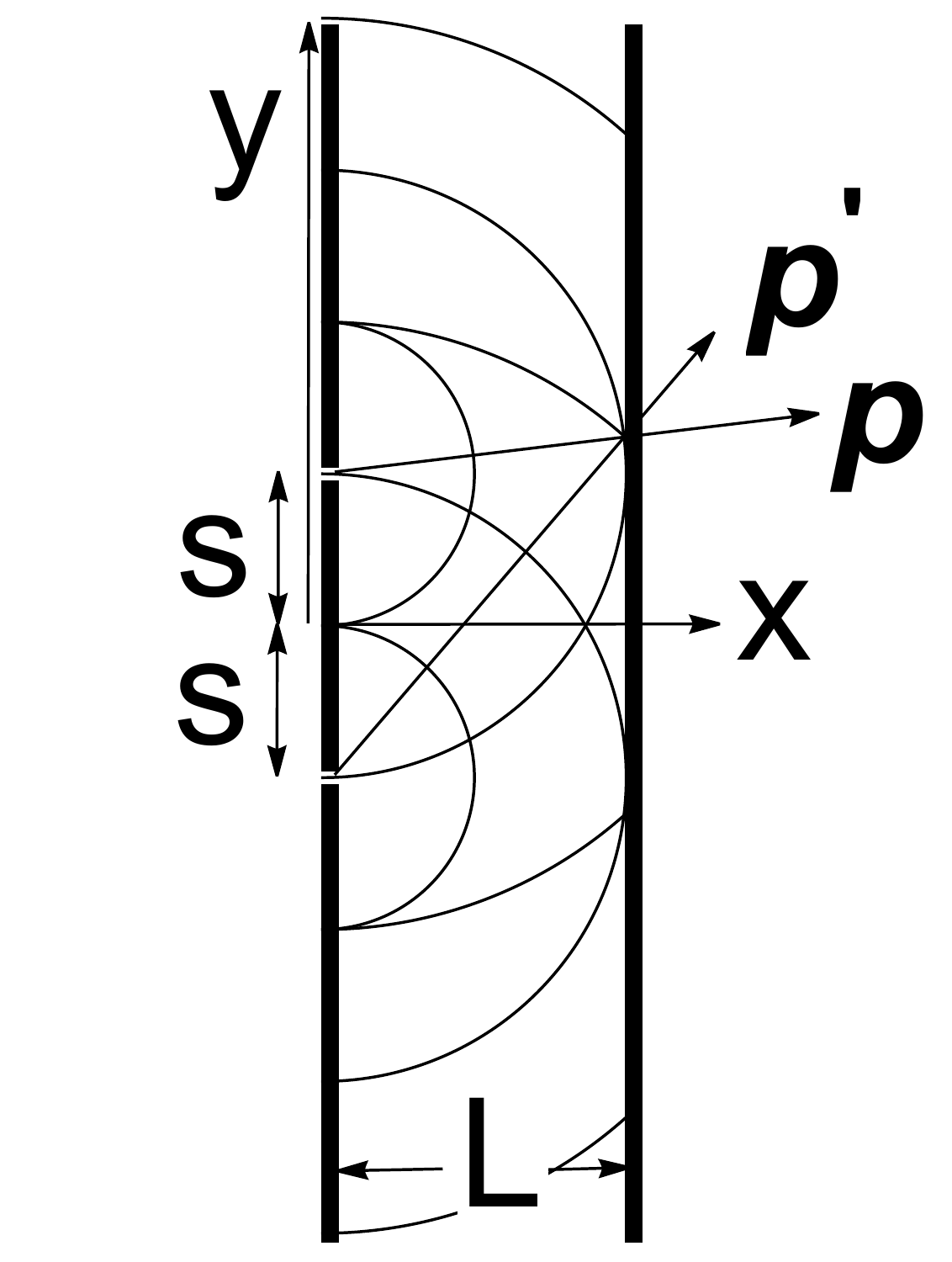}
  \caption{Setup for a double slit experiment}
  \label{fig:doubleslit}
\end{figure}

Consider the setup shown in figure \ref{fig:doubleslit} of two spherical waves centred at $(0,s)$ and $(0,-s)$ and a vertical detection screen at $x=L$. In a large separation approximation where we assume $L$ to be much larger than the vertical position along the screen $y$ and slit separation $2s$, we may approximate the spherical wave falling onto the screen as a plane wave $|{\bf p})$ perpendicular to the spherical wavefront. At a given point ${\bf x}=(L,y)$ on the screen there are two spherical waves arriving from each of the two slits. Hence, in this plane wave approximation the two waves ${\bf p}$ and ${\bf p}^\prime$ are differently orientated, but with $|{\bf p}|=|{\bf p}^\prime|$, as indicated in figure \ref{fig:doubleslit}. 

The quantum state describing the system at a point on the screen is therefore a pure state superposition of two non-commutative plane waves given by (\ref{pw}), $|\psi)=\frac{1}{\sqrt{2}}\left(|{\bf p})+|{\bf p}^\prime)\right)$. The corresponding pure state density matrix is $\rho=|\psi)(\psi|$. Using the POVM of equation (\ref{POVM}), the probability density of finding the particle at a point on the screen is now given by (\ref{prob1}) and explicitly reads
\begin{equation}\label{ncint}
    P({\bf x})=1+\rme^{-\phi^2|\bf{ p}|^2(1-\cos\alpha)}
    \cos{\left(\frac{1}{\hbar}\left({\bf p}^\prime-{\bf p}\right)\cdot{\bf x}-\phi^2|{\bf p}|^2\sin\alpha\right)},
\end{equation}
where $\alpha\in (0,2\arctan\frac{s}{L}]$ is the angle between ${\bf p}$ and ${\bf p}^\prime$, ${\bf x}=(L,y)$, $\phi^2\equiv\frac{\theta}{2\hbar^2}$, and 
\begin{equation}\label{mom}
\eqalign{
    {\bf p}^\prime&=\frac{|{\bf p}|}{\sqrt{L^2+(y+s)^2}}\left(L,y+s\right),
    \\
    {\bf p}&=\frac{|{\bf p}|}{\sqrt{L^2+(y-s)^2}}\left(L,y-s\right).
}\end{equation}
Note that the total probability is not normalised to one. For convenience, the normalisation in (\ref{pw}) was chosen so that the probability density is normalised to one in the absence of the interference term. 

Setting $\theta=0$ yields the standard commutative result, but with no modulating function as we neglected the widths of the slits and the y-dependence of the amplitude of the spherical waves under the assumption of large separation. 
This also results in the pattern evolving at high momentum to a constant rather than the anticipated two peaks. If the y-dependence of the amplitude of the spherical waves were to be taken into account, the pattern would indeed evolve into two peaks at high momentum. 

One also notes that, in contrast to the commutative case, the non-commutative result (\ref{ncint}) is not symmetric under the transformation $y\leftrightarrow -y$ due to the presence of the $\theta$ dependent term in the argument of the cosine in (\ref{ncint}), as can be explicitly verified using (\ref{mom}). As $\theta$ is presumably very small, typically of the order of the Planck length squared, this would be a very small effect. Furthermore, this asymmetry is unique to the non-commutative plane and is expected to be absent in three dimensional fuzzy space where full rotational symmetry is restored \cite{scholtz2}. It is therefore most likely not an observable physical phenomenon. 
 
Remarkably however, we do observe a Gaussian suppression of the interference term at high momentum, which arises from the off-diagonal terms of the density matrix. This would imply that the quantum properties, i.e those of quantum superposition of the particles' paths, disappear at higher momentum. We see that at any finite distance in the vertical direction, the interference term becomes suppressed when $\phi^2|{\bf p}|^2\gg 1$. Assuming $\theta$ to be of the order of the Planck length squared, this implies $|{\bf p}|\gg\phi^{-1}= \hbar\sqrt{\frac{2}{\theta}}\sim10$ kg.m.s$^{-1}$, which for microscopic particles with masses of the order of the proton mass implies speeds far in access of the speed of light for this suppression to become effective. Therefore, at non-relativistic speeds where our formalism applies, one does not expect microscopic particles to exhibit any classical behaviour. 

The non-commutative double slit experiment would seem to closely recreate what is expected from the normal commutative case for individual particles where such a description is appropriate. 
This is a reassuring result.

Considering the discussion above, that high momentum measurements are needed to probe the right-hand sector of a system state and distinguish between a pure and mixed system density matrix, it might seem contradictory that a high momentum state would suppress quantum effects. However, it should be kept in mind that the double slit experiment involves a position measurement only and, as already discussed, cannot resolve quantum correlations in the density matrix, even at high momentum.  To probe these correlations, the measurement of momentum dependent observables are required, but this cannot be done without disturbing the position measurement. The suppression of quantum interference in high momentum states is then caused by the coupling between the left-hand sector and right-hand sector, the equivalent of the environment. The adjoint nature of the momentum operator, and by extension the free Hamiltonian which causes this coupling, implies it is more efficient at high momentum and as such so is the decoherence that results.

Although the above result is insightful, it is more interesting to understand its implication for the macroscopic limit of a large number of particles where one would expect the quantum-to-classical transition to occur. We'll return to this discussion of the macroscopic limit in section \ref{sec:macroLimit}.

 Traditionally, the next step in the double split experiment is the introduction of a which path measurement into the description of the experiment. This would result in the interference pattern giving way due to the ``reduction'' of the particle's superposition path state to a single path through the two slits. We can model such a measurement as a von Neumann measurement of the form investigated next, in section \ref{sec:sternGerlach}.

\section{ von Neumann Measurement}\label{sec:sternGerlach}
A von Neumann measurement is characterised by the measurement of a ``pointer'' variable of an apparatus after it has coupled to some property of the system one is interested in. By measuring the pointer variable, one then indirectly gains insight into the nature of the system's property one wishes to measure. von Neumann style measurements have conceptual problems during the quantum to classical transition when the pointer variable is that of a typically macroscopic classical nature, such as the centre of mass of the dial on a gauge\cite{Neumann}. When the variable being measured is taken to be some quantum property of the system, such as spin, which is allowed to exist in a superposition of states, the entanglement of the system and macroscopic measuring device during the measurement process implies that the measuring device should also exist in a superposition, counter to our typical experience of macroscopic objects.

We investigate these effects here during a von Neumann measurement of the spin of a particle.  The measurement process consists of a simplified Stern–Gerlach type interaction in which a pointer is coupled to the spin of the particle through a spin-momentum coupling, described by the simplified Hamiltonian
\begin{equation}\label{SternGerlachHamiltonian}
    \hat{H}\left(t\right)=\frac{\hat{P}\hat{P}^{\ddagger}}{2m}+a\left(t\right) \hat{S} (\hat{P}^{\ddagger}+\hat{P})+H_{\rm spin},
\end{equation}
with $\hat{S}|s\>=\frac{s}{2}|s\>$, $s=\pm1$, the z-spin operator, $\hat{P}$ the centre of mass momentum operator of the pointer and $m$ its mass. In the absence of a magnetic field, as assumed here, the spin Hamiltonian is simply $H_{\rm spin}=\epsilon \hat{\bf S}^2$ with $\epsilon$ some energy scale.  Since the latter is just a constant, the dynamics are trivial and we suppress it from here on, focusing on the first two terms of the Hamiltonian only.  Furthermore $a\left(t\right)$ is a time dependent coupling function with units $[L][T^{-1}]$ and the property $a\left(t\right)\to 0$ as $t\to\infty$. In this setup, the spin of the particle is taken as the quantum property being investigated, while its state is deduced from the final position of the pointer. 

Since $\left[\hat{H}\left(t\right),\hat{H}\left(t'\right)\right]=0,\,\forall t,t'$, we can write the time evolution operator as
\begin{equation}
    \hat{U}\left(t\right)=\rme^{-\frac{i}{\hbar}\int_{0}^t \rmd{t'}\hat{H}\left(t'\right)}
\end{equation}
where we have chosen the initial time as $t_0=0$.

The pointer begins in a minimal uncertainty position state in the left-hand sector i.e. it is localized at position $z$, but the lack of knowledge about the right-hand side leaves it in a mixed Gibbs state centred on zero momentum.  
Making use of the expectation value of the Hamiltonian, $(z,w|\hat{H}|z,w)=\frac{1}{\epsilon}\abs{z-w}^2\equiv \frac{1}{\epsilon}\abs{v}^2$, with $\epsilon=\frac{m\theta}{\hbar^2}$, we subsequently define the initial density matrix as  
\begin{equation}\label{denMatrix1}
    \rho=\frac{\beta}{\pi\epsilon}\int \rmd\conj{v}\,\rmd v\: \rme^{-\frac{\beta}{\epsilon}\abs{v}^2}|\sigma\>|z,z+v)(z,z+v|\<\sigma|,
\end{equation}
where $|\sigma\>=\sum_{s=\pm} C_s|s\>$ is the normalized spin state of the particle and $\beta=\frac{1}{kT}$ with $T$ the temperature of the Gibbs state.

We then make use of the resolution of the identity in the form
\begin{equation}
    \frac{\theta}{2\pi\hbar^2}\int\rmd \conj{q}\,\rmd q |q)(q|=\mathbbm{1}_q
\end{equation}
to  transition to the momentum basis where we allow the density matrix to evolve unitarily according to the Hamiltonian given above in equation (\ref{SternGerlachHamiltonian}). This then gives a time dependent density matrix of the form
\begin{equation}\label{densmom}\eqalign{
    \rho(t) 
    =& 
    \frac{\phi^4\beta}{\pi^3 \epsilon}
    \sum_{s,s'=\pm1}{C}_s \conj{C}_{s'}
    \int\abs{\rmd v}^2\abs{\rmd q}^2\abs{\rmd k}^2
    \\&\times
    \rme^{-\frac{\beta+\epsilon}{\epsilon}\abs{v}^2}
    \rme^{-(\frac{\phi^2}{2}+i\frac{t}{2m\hbar})\abs{q}^2}
    \rme^{-(\frac{\phi^2}{2}-i\frac{t}{2m\hbar})\abs{k}^2}
    \rme^{-i\phi q\conj{v}}
    \rme^{i\phi\conj{k}v}
    \\&\times
    \rme^{-i\phi q(\conj{z}+s\tau\left(t\right))}
    \rme^{i\phi k(\conj{z}+s'\tau\left(t\right)) )}
    \rme^{-i\phi\conj{q}(z+s\tau\left(t\right)) )}
    \rme^{i\phi\conj{k}(z+s'\tau\left(t\right) ))}
    \\&\times
    |q)(k| \otimes |s\>\<s'|,
}\end{equation}
where $\tau\left(t\right) = \frac{1}{\sqrt{2\theta }}\int_{0}^{t}dt'a\left(t'\right)$ is a dimensionless quantity and $\phi\equiv\frac{1}{\hbar}\sqrt{\frac{\theta}{2}}$ has the dimension of an inverse momentum. We have also made use of the shorthand notation $\abs{\rmd v}^2=\rmd\conj{v}\rmd v$ and the momentum eigenstates are given by their plane wave representation in equation (\ref{pw}).

This density matrix is an operator on the total Hilbert space $\mathcal{H}_q\otimes\mathcal{H}_s$, where $\mathcal{H}_q$ is isomorphic to $\mathcal{H}_c\otimes\mathcal{H}^*_c$. As the measurement of the pointer variable is a position measurement, our previous considerations apply, and we have no access to the right-hand dual space sector. Following the analogy with environmental decoherence, we can perform a partial trace over the inaccessible sector. This involves representing $\rho\left(t\right)$ in the $|z,n)$ basis by inserting the resolution of the identity on both sides of $\rho\left(t\right)$ as given by equation (\ref{q_identity}) and then tracing over the states $|n\>\in\mathcal{H}_c^*$. After evaluating the $v$ and momentum integrals, this results in the reduced density matrix on the configuration space
\begin{equation}\label{reddensity}\eqalign{
    \rho_c(t) &= \frac{F(t)}{\pi^2}
    \sum_{s,s'=\pm1}{C}_s \conj{C}_{s'}\rme^{-(1-F(t))(s-s')^2\tau^2\left(t\right) \frac{m^2 \theta^2}{\hbar^2t^2}}
    \\&
    \times \int\abs{\rmd w}^2\abs{\rmd w'}^2
    \rme^{-\frac{1}{2}\abs{w-w'}^2} \rme^{\frac{1}{2}(\conj{w}w'-\conj{w}'w)} 
    \\&
    \times\rme^{-F(t)(\conj{w}-\conj{z}-s'\tau\left(t\right))(w'-z-s\tau\left(t\right))}
    \\&
    \times\rme^{-i(1-F(t))(s-s')\tau\left(t\right) \frac{m\theta}{\hbar t} (\conj{w}-\conj{z}+w'-z)} |w\>\<w'| \otimes |s\>\<s'|,
}\end{equation}
with
\begin{equation}\label{F(t)}
    F(t)=\frac{\beta m^2 \theta^2}{t^2(\beta\hbar^2+m\theta)+\beta m^2 \theta^2}.
\end{equation}

This is a two-dimensional matrix in spin space, $\mathcal{H}_s$, whose elements are operators acting on the configuration space ${\cal H}_c$. The diagonal entries of the spin space matrix, where $s'=s$, are then
\begin{equation}\label{diag}\eqalign{
{\rho_c}_{s,s}(t)=&\frac{F(t)}{\pi^2}
    \abs{C}_s^2
    \int\abs{\rmd w}^2\abs{\rmd w'}^2
    \;\rme^{-\frac{1}{2}\abs{w-w'}^2} \rme^{\frac{1}{2}(\conj{w}w'-\conj{w}'w)}
    \\&
    \times \rme^{-F(t)(\conj{w}-\conj{z}-s\tau\left(t\right))(w'-z-s\tau\left(t\right))} |w\>\<w'|,
}\end{equation}
and the off-diagonal entries, where $s'=-s$, are
\begin{equation}\label{offdiagdensity}\eqalign{
    {\rho_c}_{s,-s}(t) &= \frac{F(t)}{\pi^2}
    {C}_s \conj{C}_{-s}\rme^{-4\tau^2(t)(1-F(t))\frac{m^2 \theta^2}{\hbar^2t^2}}
    \\&
    \times \int\abs{\rmd w}^2\abs{\rmd w'}^2
    \rme^{-\frac{1}{2}\abs{w-w'}^2} \rme^{\frac{1}{2}(\conj{w}w'-\conj{w}'w)} 
    \\&
    \times\rme^{-F(t)(\conj{w}-\conj{z}+s\tau\left(t\right))(w'-z-s\tau\left(t\right))}
    \\&
    \times\rme^{-2i(1-F(t))s\tau\left(t\right) \frac{m\theta}{\hbar t} (\conj{w}-\conj{z}+w'-z)} |w\>\<w'|,
}\end{equation}
with ${\rho_c}_{ss^\prime}^\dagger={\rho_c}_{s^\prime s}$ as required.  

The probability of finding the pointer at position $r$ in dimensionless complex coordinates before the particle is observed, is simply the trace of the reduced density matrix with the operator $|r\rangle\langle r|\otimes \mathbbm{1}_s$, which yields
\begin{equation}\label{proby}
    P\left(r,\conj{r}\right)=F(t)\sum_s |C_s|^2 \rme^{-F(t)\abs{r-z-s \tau\left(t\right)}^2},
\end{equation}
with $\int\frac{|dr|^2}{\pi} P\left(r,\conj{r}\right)=1$. 

Note that in equation (\ref{offdiagdensity}) and (\ref{diag}), the factor $\rme^{\frac{1}{2}(\conj{w}w'-\conj{w}'w)}$ is just a phase, while the Gaussian factor $\rme^{-\frac{1}{2}\abs{w-w'}^2}$ implies a strong diagonalization of these operators on length scales $|w-w^\prime|\sim 1$ or, for dimensionful coordinates $x$, $|x-x^\prime|\sim\sqrt{\theta}$.  If the off-diagonal spin blocks in (\ref{offdiagdensity}) were also to vanish, this density matrix would appear as a diagonal, statistically mixed state to any position measurement not sensitive to these length scales. This suppression and its implied suppression of interference would indicate the emergence of an apparently mixed state density matrix.
 
We can track the long time behaviour of the off-diagonal blocks of the spin space density matrix by computing the trace norm of these operators,
\begin{equation}\label{trnorm}\fl\eqalign{
    ||{\rho_c}_{-+}\left(t\right)||^2
    &={\rm tr}\left({\rho_c}_{-+}^\dagger\left(t\right){\rho_c}_{-+}\left(t\right)\right)
    \\&
    =|C_+|^2|C_-|^2\frac{F(t)}{2-F(t)}\exp{-\frac{4 \tau^2\left(t\right) \left(F(t) \left(\hbar^2 t^2+2 \theta ^2 m^2\right)-2 \theta ^2 m^2\right)}{\hbar^2 t^2 (F(t)-2)}}.
}\end{equation}
It is useful to compare this with the trace norm of the diagonal spin blocks, which is found to be
\begin{equation}
\label{trnormd}
||{\rho_c}_{ss}||^2=|C_{ss}|^4\frac{F(t)}{2-F(t)}.
\end{equation}
Apart from the $C_{ss^\prime}$ factors, the only, and critical, difference between (\ref{trnorm}) and (\ref{trnormd}) is the exponential factor in (\ref{trnorm}). This implies a suppression of the off-diagonal spin blocks relative to the diagonal ones.  We return to the implications of this in our discussion below.  

For the purpose of benchmarking, let us compare these results with the corresponding commutative ones. This requires we take the zero temperature limit, where the initial density matrix (\ref{denMatrix1}) is just the pure state density matrix $\rho(0)=|\sigma\rangle|z,z)(z,z|\langle\sigma |$. In this limit, $F(t)$ reduces to
\begin{equation}\label{F(t)zero}
    F_0(t)=\frac{m^2 \theta^2}{t^2\hbar^2+m^2 \theta^2}
\end{equation}
and, after restoration of dimensions, equations (\ref{proby}), (\ref{trnorm}) and (\ref{trnormd}) reduce to
\begin{eqnarray}\label{probd}
    P\left(r,\conj{r}\right)=\left(\frac{m^2 \theta^2}{t^2\hbar^2+m^2 \theta^2}\right)\sum_s |C_s|^2 \rme^{-\frac{m^2 \theta}{t^2\hbar^2+m^2 \theta^2}| r-z-s \tau\left(t\right)|^2},
\\\label{trnormOffDiag}
    ||{\rho_c}_{-+}(t)||^2=|C_+|^2|C_-|^2\frac{\theta ^2 m^2 }{2 \hbar^2 t^2+\theta ^2 m^2}\exp{-\frac{4 \theta  m^2 \tau ^2\left(t\right)}{2 \hbar^2 t^2+\theta ^2 m^2}},\\
    \label{trnormdiag}
     ||{\rho_c}_{ss}(t)||^2=|C_{ss}|^4\frac{\theta ^2 m^2 }{2 \hbar^2 t^2+\theta ^2 m^2}
\end{eqnarray}
where $r$ and $z$ now denote the dimensionful complex coordinates $\frac{1}{\sqrt{2}}\left(x+i y\right)$, and $\tau\left(t\right)$ is now the dimensionful quantity $\tau\left(t\right)=\frac{1}{\sqrt{2}}\int_0^ta\left(t'\right)dt'$ with units of a length.

The commutative analog of the initial state density matrix would be the pure state density matrix $\rho(0)=|\sigma\rangle|z\rangle\langle z|\langle\sigma |$ where $z=\frac{1}{\sqrt{2}}\left(x+iy\right)$ are the complex dimensionful coordinates.  Evolving this density matrix with the commutative analog of the Hamiltonian (\ref{SternGerlachHamiltonian}), the time dependent commutative density matrix in position basis reads
\begin{equation}\label{comdens}\eqalign{
    \rho_{\rm com}\left(t\right)&=\frac{m^2}{\pi^2\hbar^2t^2} \sum_{s,s'=\pm1}{C}_s \conj{C}_{s'}\int |\rmd w|^2 |\rmd w'|^2
    \\&\times 
    \rme^{\frac{im}{t\hbar}|w-z-\tau s|^2} \rme^{-\frac{im}{t\hbar}|w'-z-\tau s'|^2}|w\rangle\langle w'|\otimes|s\rangle\langle s'|.
}\end{equation}
Here $z$ and $\tau$ have the same meaning as above.  As before we can easily identify the diagonal and off-diagonal operators in spin space.

Let us compare the commutative density matrix (\ref{comdens}) with the dimensionful forms of the non-commutative density matrix (\ref{diag}) and (\ref{offdiagdensity}).  The first observation we make is that, in contrast to the commutative density matrix, the off-diagonal spin blocks of the non-commutative density matrix are suppressed.   Indeed, from (\ref{trnormOffDiag}) and (\ref{trnormdiag}) we note that the off-diagonal spin blocks are suppressed relative to the diagonal ones by the exponential factor
\begin{equation}\label{supp}
 f\left(t\right)\equiv   \exp{-\tau^2(t)\frac{4m^2\theta}{2t^2\hbar^2+m^2\theta^2}},
\end{equation}
which depends on the mass of the pointer, $m$, and the nature of the coupling, $\tau(t)=\frac{1}{\sqrt{2}}\int_0^t a(t')\rmd t'$. Normally the measurement is performed over a finite time, which we denote by $T$, and for simplicity we take the coupling to be constant in this period, which gives 
\begin{equation}\label{tau}
    \tau(t)=\left\{\eqalign{
        at,\quad t<T,\\
        aT,\quad t\ge T,
}\right.\end{equation}
with irrelevant constants absorbed.  

There  are several time scales that we have to consider carefully.  The first is $\frac{m\theta}{\sqrt{2}\hbar}$, which is a very short time scale, even for macroscopic mass scales if we assume $\sqrt{\theta}$ to be of the order of the Planck length.  Since this time scale is so short, we can safely assume $\frac{m\theta}{\sqrt{2}\hbar}\ll T$.  At time scales $t\ll\frac{m\theta}{\sqrt{2}\hbar}\ll T$, the exponential in (\ref{supp}) reduces to $ \exp{-\frac{4a^2t^2}{\theta}}$, which, due to the smallness of $\theta$, leads to a rapid Gaussian suppression of the spin off-diagonal blocks with time. At time scales $\frac{m\theta}{\sqrt{2}\hbar}\ll t < T$ the exponential reduces to $\exp{-\frac{2a^2m^2\theta}{\hbar^2}}$, which is constant in time. The comparable sizes of $\theta$ and $\hbar^2$ means that this suppression term will be dominant at large mass scales and/or strong coupling, but negligible at microscopic mass scales and/or very weak coupling. At time scales $T < t$ the exponential reduces to  $\exp{-\frac{2a^2T^2m^2\theta}{t^2\hbar^2}}$, which describes the suppression of the off-diagonal spin blocks weakening and eventually disappearing at which point quantum interference is restored. This is to be expected as the pointer and system evolve as separate quantum systems after the interaction has been turned off, as compared to during the measurement process when the off-diagonal elements are suppressed.

\begin{figure}
    \centering
  \includegraphics[width=0.8\textwidth]{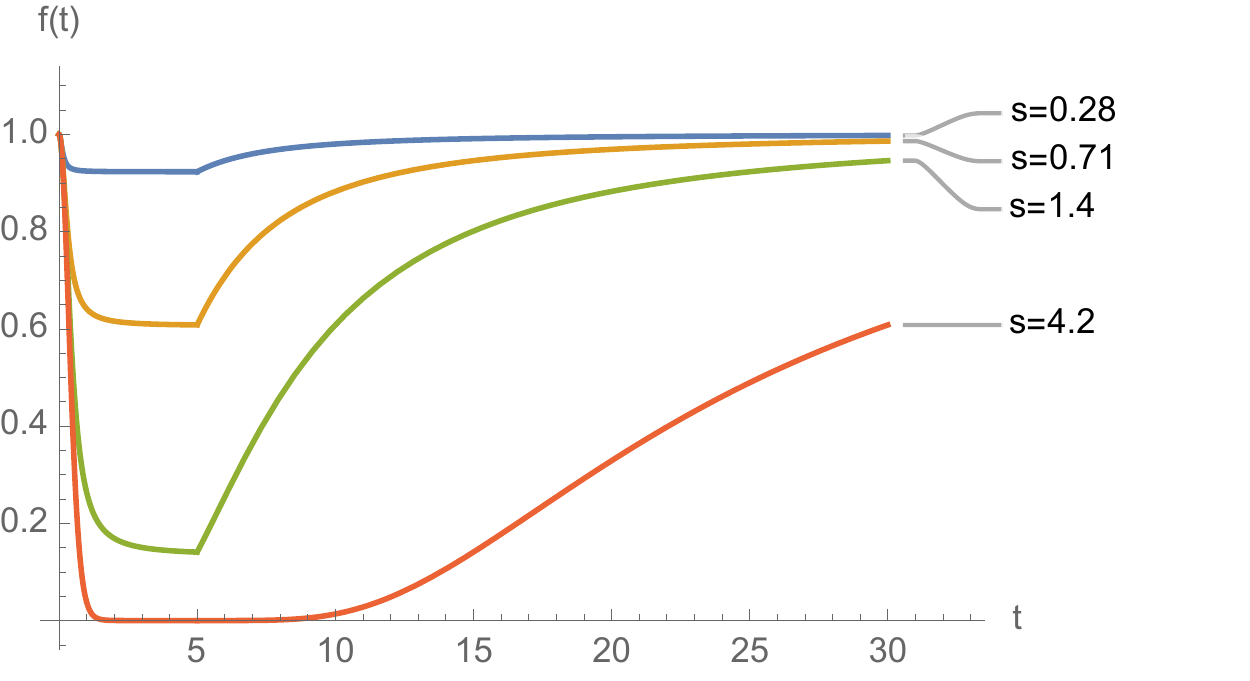}
  \caption{Behaviour of the exponential suppression of off-diagonal spin blocks as a function of time for the indicated four different values of the suppression factor $s\equiv\frac{\sqrt{2\theta}am}{\hbar}$.  Here $T=5$ and the short time scale $\frac{m\theta}{\sqrt{2}\hbar}=0.14$, $0.35$, $0.71$ and $2.12$, respectively.}
  \label{fig:supp} 
\end{figure}

If during the timescale $\frac{m\theta}{\sqrt{2}\hbar}\ll t < T$ we require that the suppression be total, then the condition $s\equiv\frac{\sqrt{2\theta}am}{\hbar}\gg 1$ must be imposed. This implies lower limits on $\theta$, the coupling constant and mass that will generally only be fulfilled on macroscopic scales.  If this condition is not satisfied, quantum interference will only be partially suppressed. This can be seen in figure \ref{fig:supp}. The four curves show the behaviour of the exponential in (\ref{supp}) for four different values of the suppression factor $\rme^{-s^2}$ with $s\equiv\frac{\sqrt{2\theta}am}{\hbar}$, corresponding to arbitrary choices of the parameters.  We note that it is only when the condition $s\gg1$ is satisfied that complete suppression occurs. 
We also note that the time scale $sT=\frac{\sqrt{2\theta}amT}{\hbar}$ at which interference returns is much later. Due to the comparable sizes of $\sqrt{\theta}$ and $\hbar$, this time scale may be macroscopic and the suppression of the off-diagonal spin blocks may occur on macroscopic time scales.

Finally we comment on the time dependence of the pre-factor occurring in (\ref{trnormOffDiag}) and (\ref{trnormdiag}).  As this is a global factor multiplying the reduced density matrix, we ignored it in our discussion above that considered the relative suppression between the diagonal and off-diagonal spin blocks.  We note that for $t>T$ this factor behaves as $\frac{\theta^2m^2}{2\hbar^2t^2}$, which is consistent with the broadening of the peaks of the probability density in (\ref{probd}) and required to ensure the normalization of the probability.  This behaviour is further expanded on below.

We have established that under appropriate conditions the off-diagonal spin blocks in the non-commutative density matrix are absent. We now focus on the diagonal spin blocks.  Considering the zero temperature limit of the general spin space diagonal matrix element (\ref{diag}) and restoring dimensions, this reads
\begin{equation}\eqalign{
    {\rho_c}_{s,s}(t)=&\frac{m^2}{\pi^2\left(t^2\hbar^2+m^2 \theta^2\right)}
    \abs{C}_s^2
    \int\abs{\rmd w}^2\abs{\rmd w'}^2
    \;\rme^{-\frac{1}{2\theta}\abs{w-w'}^2} \rme^{\frac{1}{2\theta}(\conj{w}w'-\conj{w}'w)}
    \\&
    \times \rme^{-\frac{m^2 \theta}{t^2\hbar^2+m^2 \theta^2}(\conj{w}-\conj{z}-s\tau(t))(w'-z-s\tau(t))} |w\>\<w'|.
}\end{equation}
From this, it is clear that the spin diagonal blocks of the non-commutative density matrix are localized on the diagonal to a length scale of $\sqrt{\theta}$. This corresponds to the variance of the minimum uncertainty states, $|z\>$, making this a mixed density matrix of the closest thing the non-commutative program has to position eigenstates. In fact, in the $\theta\to 0$ limit, it becomes diagonal with precisely the same constant diagonal matrix elements as the commutative density matrix $\rho_{com}$. Keeping in mind that under appropriate conditions the off-diagonal spin blocks of the non-commutative density matrix are suppressed and that the diagonal spin blocks are localized on the diagonal to a length scale of  $\sqrt{\theta}$, we conclude that the full non-commutative density matrix assumes, under these conditions, a diagonal form on the length scale $\sqrt{\theta}$ and that any observer that cannot measure down to this length scale will interpret it as a mixed state.  As the non-commutative and commutative density matrices coincide to lowest order in $\theta$ on the diagonal, they will yield the same statistics for position measurements, but have different correlations.  

Critically, the non-commutative density matrix is a mixed state due to the partial trace over ${\cal H}_c^*$, while the commutative one is a pure density matrix as it involves the unitary evolution of a pure state, which accounts for the fact that its off-diagonal matrix elements are not suppressed as in the non-commutative case.

With the properties of the non-commutative density matrix established, we have the appropriate context to interpret the probability density (\ref{probd}).  As we are interested in the final position of the pointer, we consider the limit $\frac{m\theta}{\sqrt{2}\hbar}\ll T < t$.  Making use of (\ref{tau}),  we note that the probability is peaked at two points $r_+=z+\tau\left(t\right)=z+aT$ and $r_-=z-\tau\left(t\right)=z-aT$ with weights $|C_+|^2$ and $|C_-|^2$, respectively.  As expected, this indicates that the pointer will be found in these two positions with the corresponding probabilities.  The mixed nature of the non-commutative density matrix, as established above, implies that the observer will interpret this as the pointer being at either the position $r_+$ or $r_-$ with the indicated probabilities. 

At this time scale the probability is peaked at these two points on a length scale $\frac{t\hbar}{m\sqrt{\theta}}$, implying a broadening with time as was noted before.  At large mass scales this will generally be a short length scale leading to a sharply peaked probability distribution with the broadening only emerging at time scales $t\sim \frac{m\sqrt{\theta}}{\hbar}$, which, due to $\sqrt{\theta}$ and $\hbar$ being of similar order, could be a macroscopic time scale.

From the above discussion it is clear that the size or mass of the pointer plays a critical role and that the suppression of the off-diagonal spin blocks of the density matrix will be minute for a microscopic pointer.  At the same time the mass dependence suggests that the situation may change in the macroscopic limit, which requires us to make the dependence on the size of the pointer explicit, which is the topic of the next section.

\section{\label{sec:macroLimit}The Macroscopic Limit}
To better understand this transition from quantum to classical behaviour, let us first recall the situation in the commutative case. Typically, for a macroscopic object built from microscopic particles, the inter-particle potential only depends on the relative distance between the particles and the Hamiltonian is therefore translationally invariant. This results in the complete decoupling of the centre of mass and relative coordinates, with the centre of mass dynamics becoming that of a free particle with mass equal to the total mass of the object. This presents somewhat of a dilemma for commutative quantum mechanics as without the introduction of environmental decoherence \cite{legget} one would expect the centre of mass of this macroscopic object to behave quantum mechanically.

Applying this argument to the double split experiment, one would expect interference due to the quantum superposition of the object's path through the two slits, contradictory to its classical macroscopic nature \cite{Neumann}. Remarkably, in the non-commutative case, there is no such contradiction. The suppression of the quantum behaviour happens naturally within this framework, regardless of the presence of an environment.

Consider a macroscopic object which we take for simplicity to be composed of $N$ particles with the same mass, $m$. We denote the non-commutative coordinates of the particles by $\{\hat{\vb{x}}_{i}\}_{i=1}^N$. The coordinates of different particles commute, $[\hat{\vb{x}}^{(k)}_i,\hat{\vb{x}}^{(l)}_j]= i\theta\delta_{ij}\epsilon_{k,l}$, $k,l=1,2$. The Hilbert spaces ${\cal H}_c$ and ${\cal H}_q$ are correspondingly generalised to tensor product spaces. Introducing the centre of mass coordinates, $\hat{\vb{x}}_0=\frac{1}{N}\sum_i^N \hat{\vb{x}}_{i}$, one notices that $[\hat{\vb{x}}_0^{(i)},\hat{\vb{x}}_0^{(j)}]=i\frac{\theta}{N}\epsilon_{i,j}\equiv i\tilde\theta\epsilon_{i,j}$. Thus, the centre of mass coordinates are non-commutative coordinates with a non-commutative parameter $\tilde\theta$.  As before the coordinates are promoted to operators acting on the quantum Hilbert space through left multiplication and we again use capitals to distinguish them. One can then introduce the total momentum operator $\hat{\vb{P}}_0=\sum_i^N\hat{\vb{P}}_{i}$ as an operator acting on the quantum Hilbert space. It can be verified from (\ref{momadj}) that its action on a generic element $|\psi)$ of the quantum Hilbert space is given by 
\begin{equation}
    \hat{P}_0^{(i)}|\psi)=|\frac{\hbar}{\tilde\theta}\epsilon_{i,j}[\hat{\vb{x}}_0^{(j)},\psi]),
\end{equation}
which is consistent with the commutation relations of the centre of mass coordinates.

Let us also introduce the relative position and momentum coordinates, $\hat{\vb{R}}_{i}=\hat{\vb{X}}_{i}-\hat{\vb{X}}_0$ and $\hat{\vb{Q}}_{i}=\hat{\vb{P}}_{i}-\hat{\vb{P}}_0$, which are interdependent and satisfy $\sum_i^N\hat{\vb{R}}_{i}=0$ and $\sum_i^N\hat{\vb{Q}}_{i}=0$ respectively. The centre of mass coordinates $\hat{\vb{X}}_0$ and momentum $\hat{\vb{P}}_0$ satisfy standard canonical commutation relations, while both commute with all relative coordinates and momenta. 

The Hamiltonian for both the double split experiment and that used in the von Neumann measurement in equation (\ref{SternGerlachHamiltonian}) can be rewritten as
\begin{equation}\eqalign{
    H&=\frac{\hat{\vb{P}}_0^2}{2M}+V_0+H_{\rm int}
    \\
    &=H_0+H_{\rm int}.
}\end{equation}
Here $M=Nm$ is the total mass and $H_{\rm int}$ describes the internal and possible spin dynamics and only depends on the relative coordinates. The Hamiltonian is then separable, with the centre of mass Hamiltonian $H_0$ that of a point particle with mass $M$ in a non-commutative plane with effective non-commutative parameter $\tilde\theta$.

The results of (\ref{ncint}) therefore apply with $\theta$ replaced by $\tilde\theta$ and $\vb{p}$ replaced by $\vb{p}_0$.  Keeping in mind that the total momentum is extensive, i.e. $\vb{p}_0=Nm\vb{v}$ with $\vb{v}$ the velocity of the object and therefore, for a rigid object, the average velocity of the composing particles, we note that (\ref{ncint}) develops an $N$ dependence,
\begin{equation}\label{ncintmac}
    P({\bf x})=1+\rme^{-\frac{N\theta m^2}{2\hbar^2}\vb{v}^2(1-\cos\alpha)}
    \cos\left(\frac{1}{\hbar}\left({\bf p_0}^\prime-{\bf p_0}\right)\cdot{\bf x}-\frac{N\theta m^2\vb{v}^2}{2\hbar^2}\sin\alpha\right).
\end{equation}
It is then clear that the interference term gets suppressed at any finite vertically displaced point on the screen as $N$ is increased, naturally leading to what would be expected from classical behaviour. This, of course, also destroys the asymmetry of the interference pattern at macroscopic scales, also keeping in mind the earlier remarks made in this regard. This matches the behaviour of the visibility of the interference bands in interferometer experiments with $C_{70}$ atoms interacting with an increasing environment \cite{Hornberger} if the size $N$ of the system is made to include the particles of the atmosphere the atom has interacted with.

Considering (\ref{ncintmac}), we note that for the Gaussian suppression to be effective requires
\begin{equation}
    \theta\gg\frac{\hbar^2}{N m^2{\bf v}^2}.
\end{equation}
We do not know for which values of $N$ the classical transition occurs \cite{legget} and, as suggested by the above results, it is most likely a continuous transition \cite{legget}. The proponents of collapse models would suggest a transition at $N\sim10^{13}$, with some proposing values as low as $N\sim10^5$ \cite{bassi_2010}. However, these refer to three dimensional systems, and as such we choose here a conservative estimate for the cross over point to be that of the order of Avogadro's number. Making use of non-relativistic speeds with $|{\bf v}|\sim 1$ m.s$^{-1}$ and considering only the contribution of the protons and neutrons (taking their masses to be equal) to the mass of the macroscopic object, we conclude $\theta\gg10^{-40}$m$^2$. This is much greater than the generally believed value of $\theta$ as the Planck length squared, but still small enough to prevent any classical effects emerging for a single microscopic particle, which will require speeds well above the speed of light. 

Given that all observational data provide an upper bound to $\theta$, the emergence of a lower bound is an interesting and remarkable result. In this context, the interplay between microscopic and macroscopic phenomena may provide stringent tests for non-commutativity and constraints on the non-commutative parameter. It must, however, be kept in mind that the non-commutative plane is an idealisation and for comprehensive experimental comparison these results have to be generalised to three-dimensions with unbroken rotational symmetry. It will probably also be necessary to take much greater care of modulating effects.

The same argument can be applied to the von Neumann measurement where the system couples to the centre of mass momentum of the pointer with the conclusion that all the results of the previous section apply with the replacements of $\theta\rightarrow\tilde{\theta}=\frac{\theta}{N}$ and $m\rightarrow Nm$.  Here we focus on the zero temperature limit results, that of equation (\ref{probd}) and equation (\ref{trnormOffDiag}), which then reads
\begin{eqnarray}\label{probmacN}
    P\left(r,\conj{r}\right)=\left(\frac{m^2 \theta^2}{t^2\hbar^2+m^2 \theta^2}\right)\sum_s |C_s|^2 \rme^{-N\frac{m^2 \theta}{t^2\hbar^2+m^2 \theta^2}| r-z-s \tau|^2},
\\\label{offmacN}
    ||{\rho_c}_{-+}(t)||^2=|C_+|^2|C_-|^2\frac{\theta ^2 m^2 }{2 \hbar^2 t^2+\theta ^2 m^2}\exp{-\frac{4N \theta  m^2 }{2 \hbar^2 t^2+\theta ^2 m^2}\tau(t) ^2}.
\end{eqnarray}

Note that the pre-factors and thus (\ref{trnormdiag}) is unaltered and therefore not listed here again.  Applying the results from the previous section, we firstly note that the short time scale $\frac{m\theta}{\sqrt{2}\hbar}$ is unaffected and that the condition $s\equiv\frac{\sqrt{2\theta}am}{\hbar}\gg 1$ is now $\sqrt{N}s=\frac{\sqrt{2N\theta}am}{\hbar}\gg 1$, exhibiting an explicit size dependence.  This condition can clearly be satisfied for a large enough pointer size.   At time scales $\frac{m\theta}{\sqrt{2}\hbar}\ll t <T$ the exponential suppression factor of the off-diagonal spin blocks becomes $\rme^{-Ns^2}$, also exhibiting an explicit size dependence that enhances the suppression.  Indeed, we see that complete suppression of the quantum interference can be achieved for a large enough pointer size.
Additionally, the time scale $\frac{\sqrt{2\theta}amT}{\hbar}$ at which quantum interference revives now turns into $ \frac{\sqrt{2N\theta}amT}{\hbar}$, again exhibiting an explicit size dependence that emphasizes its macroscopic nature.  This time scale can clearly become arbitrarily large for a large enough system size.

Finally, from the probability density (\ref{probmacN}) at time scales $\frac{m\theta}{\sqrt{2}\hbar}\ll T < t$, the probability is peaked at the points $r_+=z+aT$ and $r_-=z-aT$ on a length scale that is now $\frac{t\hbar}{\sqrt{N\theta}m}$, which is again size dependent and can be made very short for large enough pointer size. This implies a sharply peaked probability distribution at $r_+$ and $r_-$. Similarly, the time scale on which the peaks broaden is now $t\sim\frac{m\sqrt{N\theta}}{\hbar}$ and can be made arbitrarily large for large enough pointer size.

It is difficult to make realistic estimates of the various quantities above, especially due to the unknown nature of the coupling constant $a$.  It is clear though that to reconcile these quantities with observational data will require rather stringent limits on $\theta$.  However, as already remarked, it should be kept in mind that the two-dimensional plane is an idealization and that this scenario may be quite different in a full three dimensional analysis.

\section{Discussion and conclusion}\label{sec:discussion}
This paper aimed to explore what is as yet a virtually unexplored question, namely, does the structure of space at short length scales have a macroscopic imprint?  Indeed, such a link is most likely our only hope of ever probing the structure of space at short length scales. At this point, the only way we could see to proceed was to adopt a specific paradigm for space at short length scales, which we took to be non-commutative. We then worked out the consequence of this scenario in the non-commutative plane for quantum interference phenomena, specifically in a double slit experiment and a von Neumann measurement.  

The outcome is a rather pleasant result. In both cases the partial suppression of interference phenomena tends to complete suppression at large $N$. This describes the continuous transition of the system to a scale where its quantum properties are suppressed, governed exclusively by the scale, i.e. the number of constituent particles $N$. It is a description of a non-unitary transition of the system from quantum to classical behaviour as it transitions from microscopic to macroscopic, without the need for the introduction of any new external systems to drive this process such as an environment. 

This behaviour also identifies position as a preferred basis in which this transition occurs, driven by the inability of such a measurement to access the right-hand sector of a systems density matrix. In this case, the system may undergo a non-unitary evolution to what appears as a mixed state to any observer who cannot probe length scales beyond $\sqrt{\theta}$. During such a position measurement, the right-hand sector becomes extra degrees of freedom beyond the reach of an observer.
For full access to the right-hand sector and for complete knowledge of the system state, measurements of unreasonably high momentum are required, well beyond what is operationally feasible.

It is not unreasonable to say that the non-commutative program, through the apparent suppression of interference in macroscopic objects and emergence of a preferred position basis, offers a possible natural resolution to the problem of macro-objectivity. That this program follows from arguments seeking a quantum description of gravity, a seemingly distantly related field, is fascinating. Additionally, each of these perspectives offer a lower bound on the non-commutative parameter

Naturally, the description given here is limited in scope and dimension and would need to be expanded, but the insight it offers is quite interesting and worthy of further exploration.

\section*{Acknowledgements}
This work is based on the research supported by the National Research Foundation of South Africa.

\section*{References}
\bibliographystyle{iopart-num}
\bibliography{Bibliography.bib}

\end{document}